\documentclass[sigconf]{acmart}

\usepackage{booktabs} 
\usepackage{algorithmic}
\usepackage[linesnumbered, ruled, vlined]{algorithm2e}
\usepackage{multirow}
\usepackage{tablefootnote}
\usepackage{graphicx}
\usepackage{subfigure}
\usepackage{flushend}
\usepackage{color}
\usepackage{enumitem}
\usepackage{array, makecell}
\usepackage{footmisc}
\usepackage{amsmath}
\usepackage{xcolor}

\newcommand{\hide}[1]{}
\newcommand{\vpara}[1]{\vspace{0.07in}\noindent\textbf{#1 }}
\newcommand{\model}{{ComiRec}}

\AtBeginDocument{%
  \providecommand\BibTeX{{%
    \normalfont B\kern-0.5em{\scshape i\kern-0.25em b}\kern-0.8em\TeX}}}

\copyrightyear{2020}
\acmYear{2020}
\setcopyright{acmcopyright}
\acmConference[KDD '20]{Proceedings of the 26th ACM SIGKDD Conference on Knowledge Discovery and Data Mining}{August 23--27, 2020}{Virtual Event, CA, USA}
\acmBooktitle{Proceedings of the 26th ACM SIGKDD Conference on Knowledge Discovery and Data Mining (KDD '20), August 23--27, 2020, Virtual Event, CA, USA}
\acmPrice{15.00}
\acmDOI{10.1145/3394486.3403344}
\acmISBN{978-1-4503-7998-4/20/08}

\begin{document}

\fancyhead{}

\title{Controllable Multi-Interest Framework for Recommendation}

\author[Y. Cen, J. Zhang, X. Zou, C. Zhou, H. Yang and J. Tang]{
    Yukuo Cen$^{\dagger}$, Jianwei Zhang$^{\ddagger}$, Xu Zou$^{\dagger}$, Chang Zhou$^{\ddagger}$, Hongxia Yang$^{\ddagger*}$, Jie Tang$^{\dagger*}$
}
\affiliation{
    $^\dagger$ Department of Computer Science and Technology, Tsinghua University
}
\affiliation{
    $^\ddagger$ DAMO Academy, Alibaba Group
}
\email{
  {cyk18, zoux18}@mails.tsinghua.edu.cn
}
\email{
  {zhangjianwei.zjw, ericzhou.zc, yang.yhx}@alibaba-inc.com
}
\email{
  jietang@tsinghua.edu.cn
}

\renewcommand{\authors}{Yukuo Cen, Jianwei Zhang, Xu Zou, Chang Zhou, Hongxia Yang and Jie Tang}
\renewcommand{\shortauthors}{Y. Cen et al.}

\begin{abstract}
\renewcommand{\thefootnote}{\fnsymbol{footnote}}
\footnotetext[1]{Hongxia Yang and Jie Tang are the corresponding authors.}
\renewcommand{\thefootnote}{\arabic{footnote}}

Recently, neural networks have been widely used in e-commerce recommender systems, owing to the rapid development of deep learning. 
We formalize the recommender system as a sequential recommendation problem, intending to predict the next items that the user might be interacted with. 
Recent works usually give an overall embedding from a user's behavior sequence. However, a unified user embedding cannot reflect the user's multiple interests during a period. 
In this paper, we propose a novel \textbf{\textit{co}}ntrollable \textbf{\textit{m}}ulti-\textbf{\textit{i}}nterest framework for the sequential \textbf{\textit{rec}}ommendation, called \model. Our multi-interest module captures multiple interests from user behavior sequences, which can be exploited for retrieving candidate items from the large-scale item pool. These items are then fed into an aggregation module to obtain the overall recommendation. The aggregation module leverages a controllable factor to balance the recommendation accuracy and diversity. 
We conduct experiments for the sequential recommendation on two real-world datasets, Amazon and Taobao. Experimental results demonstrate that our framework achieves significant improvements over state-of-the-art models\footnote{Code is available at \url{https://github.com/THUDM/ComiRec}}. 
Our framework has also been successfully deployed on the offline Alibaba distributed cloud platform. 
\end{abstract}



\maketitle

\section{Introduction}\label{sec:intro}

The development of e-commerce revolutionized our shopping styles in recent years. Recommender systems play a fundamental role in e-commerce companies. 
Traditional recommendation methods mainly use collaborative filtering methods~\cite{sarwar2001item,schafer2007collaborative} to predict scores between users and items. Recently, neural networks have been widely used in e-commerce recommender systems, owing to the rapid development of deep learning. Neural recommender systems generate representations for users and items and outperform traditional recommendation methods. However, due to the large-scale e-commerce users and items, it is hard to use deep models to directly give the click-through rate (CTR) prediction between each pair of users and items. Current industrial practice is to use fast K nearest neighbors (e.g., Faiss~\cite{JDH17}) to generate the candidate items and then use a deep model (e.g., xDeepFM~\cite{lian2018xdeepfm}) to integrate the attributes of users and items to optimize the business metrics such as CTR.


Some recent works leverage graph embedding methods to obtain representations for users and items, which can be used for downstream applications. For example, PinSage~\cite{ying2018graph} builds on GraphSAGE~\cite{hamilton2017inductive} and has applied graph convolutional network based methods to production-scale data with billions of nodes and edges. GATNE~\cite{cen2019representation} considers different user behavior types and leverages a heterogeneous graph embedding method to learn representations for users and items. However, this kind of method ignores the sequential information in the user behaviors and cannot capture the correlations between adjacent user behaviors. 

\begin{figure*}
    \centering
    \includegraphics[width=0.8\textwidth]{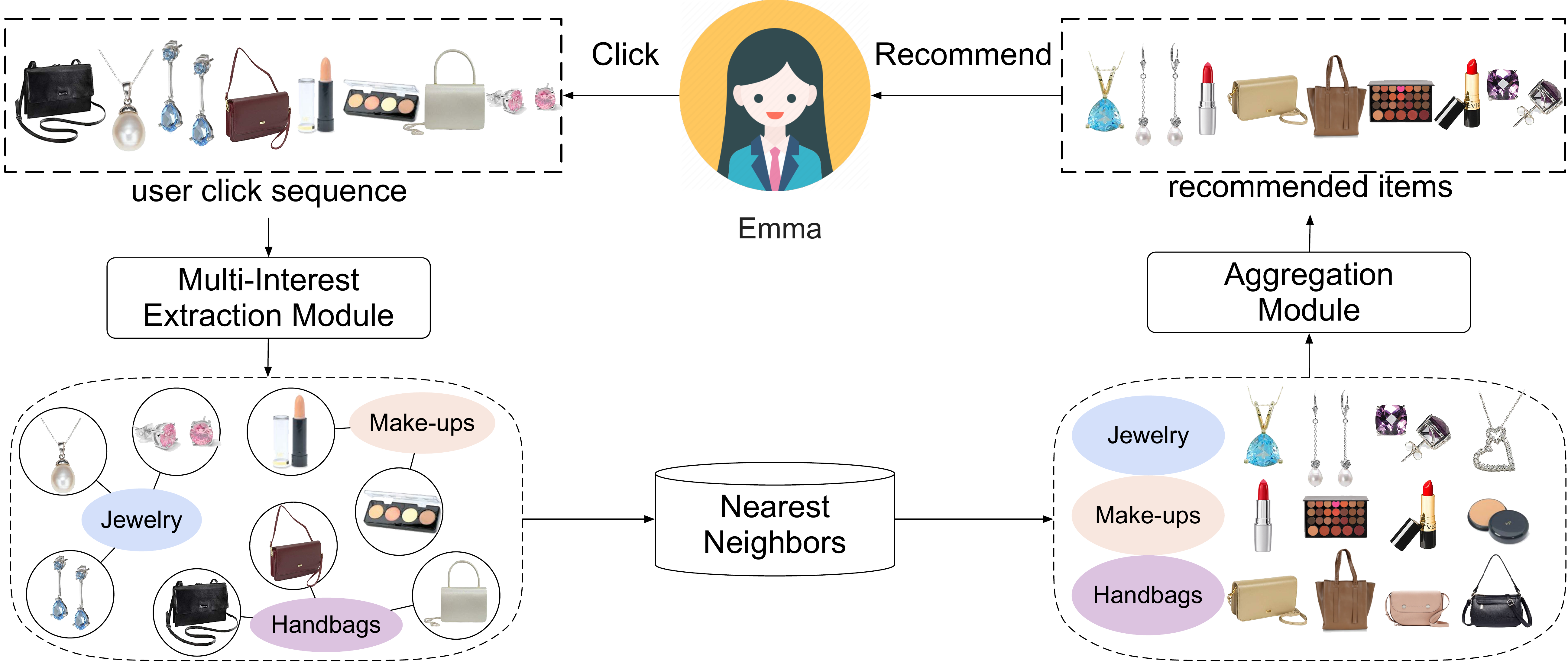}
    \caption{A motivating example of our proposed framework. An e-commerce platform user, Emma, has multiple interests including jewelry, handbags, and make-ups. Our multi-interest extraction module can capture these three interests from her click sequence. Each interest retrieves items from the large-scale item pool based on the interest embedding independently. An aggregation module combines items from different interests and outputs the overall top-N recommended items for Emma. }
    \label{fig:multiple_interest}
\end{figure*}

Recent researches~\cite{kang2018self,chen2018sequential,lv2019sdm} formalize the recommender system as a sequential recommendation problem. With a user's behavior history, the sequential recommendation task is to predict the next item he/she might be interested in. This task reflects the real-world recommendation situation. Many recent models can give an overall embedding for each user from his/her behavior sequence. However, a unified user embedding is hard to represent multiple interests. For example, in Figure~\ref{fig:multiple_interest}, the click sequence shows three different interests of Emma. As a modern girl, Emma is interested in jewelry, handbags, and make-ups. Therefore, she may click items of the three categories during this period of time. 


In this paper, we propose a novel controllable multi-interest framework, called \model. Our multi-interest module can capture the multiple interests of users, which can be exploited for retrieving candidate items. Our aggregation module combines these items from different interests and outputs the overall recommendation. Figure~\ref{fig:multiple_interest} shows a motivating example of our multi-interest framework. We conduct experiments for the sequential recommendation, which is similar to our online situation. The experimental results show that our framework outperforms other state-of-the-art models. Our framework has also been successfully deployed on the Alibaba distributed cloud platform. Results on the billion-scale industrial dataset further confirm the effectiveness and efficiency of our model in practice. 

To summarize, the main contributions of this paper are:
\begin{itemize}
    \item We propose a comprehensive framework that integrates the controllability and multi-interest components in a unified recommender system. 
    \item We investigate the role of controllability on personalized systems by implementing and studying in an online recommendation scenario. 
    \item Our framework achieves state-of-the-art performance on two real-world challenging datasets for the sequential recommendation.
\end{itemize}

\section{Related Work} \label{sec:related}

In this section, we introduce the related literature about recommender systems and recommendation diversity, as well as capsule networks and the attention mechanism we used in the paper.

Collaborative filtering~\cite{sarwar2001item,schafer2007collaborative} methods have been proven successful in real-world recommender systems, which find similar users and items and make recommendations on this basis. Matrix factorizaion~\cite{koren2009matrix} is the most popular technique in classical recommender research, which maps both users and items to a joint latent factor space, such that user-item interactions are modeled as inner products in that space. Factorization Machines (FMs)~\cite{rendle2010factorization} model all interactions between variables using factorized parameters and thus can estimate interactions even in problems with huge sparsity like recommender systems.

\vpara{Neural Recommender Systems.}
Neural Collaborative Filtering (NCF)~\cite{he2017neural} uses a neural network architecture to model latent features of users and items. 
NFM~\cite{he2017nfm} seamlessly combines the linearity of FMs in modeling second-order feature interactions and the non-linearity of neural networks in modeling higher-order feature interactions.
DeepFM~\cite{guo2017deepfm} designs an end-to-end learning model that emphasizes both low-order and high-order feature interactions for CTR prediction.
xDeepFM~\cite{lian2018xdeepfm} extends DeepFM and can learn specific bounded-degree feature interactions explicitly.
Deep Matrix Factorization (DMF)~\cite{xue2017deep} uses a deep structure learning architecture to learn a common low dimensional space for the representations of users and items based on explicit ratings and non-preference implicit feedback.
DCN~\cite{wang2017deep} keeps the benefits of a deep model and introduces a novel cross network that is more efficient in learning specific bounded-degree feature interactions.
CMN~\cite{ebesu2018collaborative} uses deep architecture to unify the two classes of CF models capitalizing on the strengths of the global structure of the latent factor model and local neighborhood-based structure in a nonlinear fashion.

\vpara{Sequential Recommendation.}
The sequential recommendation is the crucial problem of recommender systems. Many recent works about recommender systems focus on this problem. FPMC~\cite{rendle2010factorizing} subsumes both a common Markov chain and the normal matrix factorization model for sequential basket data. HRM~\cite{wang2015learning} extends the FPMC model and employs a two-layer structure to construct a hybrid representation over users and items from the last transaction. GRU4Rec~\cite{hidasi2015session} first introduces an RNN-based approach to model the whole session for more accurate recommendations. DREAM~\cite{yu2016dynamic}, based on Recurrent Neural Network (RNN), learns a dynamic representation of a user for revealing the user's dynamic interests. Fossil~\cite{he2016fusing} integrates similarity-based methods with Markov Chains smoothly to make personalized sequential predictions on sparse and long-tailed datasets. TransRec~\cite{he2017translation} embeds items into a vector space where users are modeled as vectors operating on item sequences for large-scale sequential prediction. RUM~\cite{chen2018sequential} uses a memory-augmented neural network integrated with the insights of collaborative filtering for the recommendation. SASRec~\cite{kang2018self} uses a self-attention based sequential model to capture long-term semantics and uses an attention mechanism to make its predictions based on relatively few actions. DIN~\cite{zhou2018deep} designs a local activation unit to adaptively learn the representation of user interests from past behaviors with respect to a certain ad. SDM~\cite{lv2019sdm} encodes behavior sequences with a multi-head self-attention module to capture multiple types of interests and a long-short term gated fusion module to incorporate long-term preferences.


\vpara{Recommendation Diversity.}
Researchers have realized that following only the most accurate way of recommendation may not result in the best recommendation results, since the highest accuracy results tend to recommend similar items to users, yielding boring recommendation results~\cite{panniello2014comparing}. To address such problems, the diversity of the recommended items also plays a significant role~\cite{slaney2006measuring}. In terms of diversity, there is aggregated diversity~\cite{adomavicius2011improving}, which refers to the ability to recommend "long-tail items" to users. Many studies focus on improving aggregated diversity of recommendation systems  ~\cite{bag2019integrated,adomavicius2011improving,niemann2013new,qin2013promoting}. Other works focus on the diversity of items recommended to individual users, i.e., the individual diversity~\cite{adomavicius2011improving,yu2019recommendation,kalaivanan2013recommendation,di2014analysis}, which refers to the dissimilarity of items recommended to an individual user.


\vpara{Attention}
The originality of attention mechanism can be traced back to decades ago in fields of computer vision ~\cite{burt1988attention,sun2003object}. However, its popularity in various fields in machine learning comes only in recent years. It is first introduced to machine translation by ~\cite{bahdanau2014neural}, and later becomes an outbreaking method as \textit{tensor2tensor}~\cite{vaswani2017attention}. BERT ~\cite{devlin2018bert} leverages \textit{tensor2tensor} and achieves giant success in natural language processing. 
The attention mechanism is also adapted to recommender systems ~\cite{zhou2018atrank,cen2019representation} and is rather useful on real-world recommendation tasks. 

\vpara{Capsule Network.}
The concept of ``capsules'' is first proposed by~\cite{hinton2011transforming} and has become well-known since the dynamic routing method ~\cite{sabour2017dynamic} is proposed. 
MIND~\cite{li2019multi} introduces capsules into recommendation areas and uses the capsule network to capture multiple interests of e-commerce users based on dynamic routing mechanism, which is applicable for clustering past behaviors and extracting diverse interests. CARP~\cite{li2019capsule} firstly extracts the viewpoints and aspects from the user and item review documents and derives the representation of each logic unit based on its constituent viewpoint and aspect for rating prediction.

\section{Methodology}\label{sec:model}
In this section, we formulate the problem and introduce the proposed framework in detail, as well as showing the difference between our framework and representative existing methods.

\begin{table}
  \centering
  \caption{\label{tab:notation} Notations.}
  \begin{tabular}{c|p{2.55in}}
    \hline \hline
    \textbf{Notation} & \textbf{Description} \\
    \hline
    $u$ & a user \\
    $i$ & an item \\
    $e$ & an interaction \\
    $\mathcal{U}$ & the set of users \\
    $\mathcal{I}$ & the set of items \\
    $\mathcal{I}_u$ & the set of testing items of user $u$ \\
    $d$ & the dimension of user/item embeddings \\
    $K$ & the number of interest embeddings \\
    $N$ & the number of candidate items\\
    $\mathbf{V}_u$ & the matrix of interest embeddings of user $u$ \\
    $\delta(\cdot)$ & indicator function \\
    \hline \hline
  \end{tabular}
\end{table}

\subsection{Problem Formulation}
Assume we have a set of users $u\in \mathcal{U}$ and a set of items $i\in \mathcal{I}$. For each user, we have a sequence of user historical behaviors $(e^{(u)}_{1}, e^{(u)}_{2}, \cdots, e^{(u)}_{n})$, sorted by time of the occurrence. $e^{(u)}_{t}$ records the $t^{th}$ item interacted by the user. Given historical interactions, the problem of \textit{sequential recommendation} is to predict the next items that the user might be interacted with. 
Notations are summarized in Table~\ref{tab:notation}.

In practice, due to the strict requirements of latency and performance, industrial recommender systems usually consist of two stages, the matching stage and the ranking stage. The matching stage corresponds to retrieving top-N candidate items, while the ranking stage is used for sorting the candidate items by more precise scores. Our paper mainly focuses on improving the effectiveness in the matching stage. In the following parts of this section, we will introduce our controllable multi-interest framework and illustrate the significance of our framework for the \textit{sequential recommendation} problem. 

\begin{figure*}
    \centering
    \includegraphics[width=0.8\textwidth]{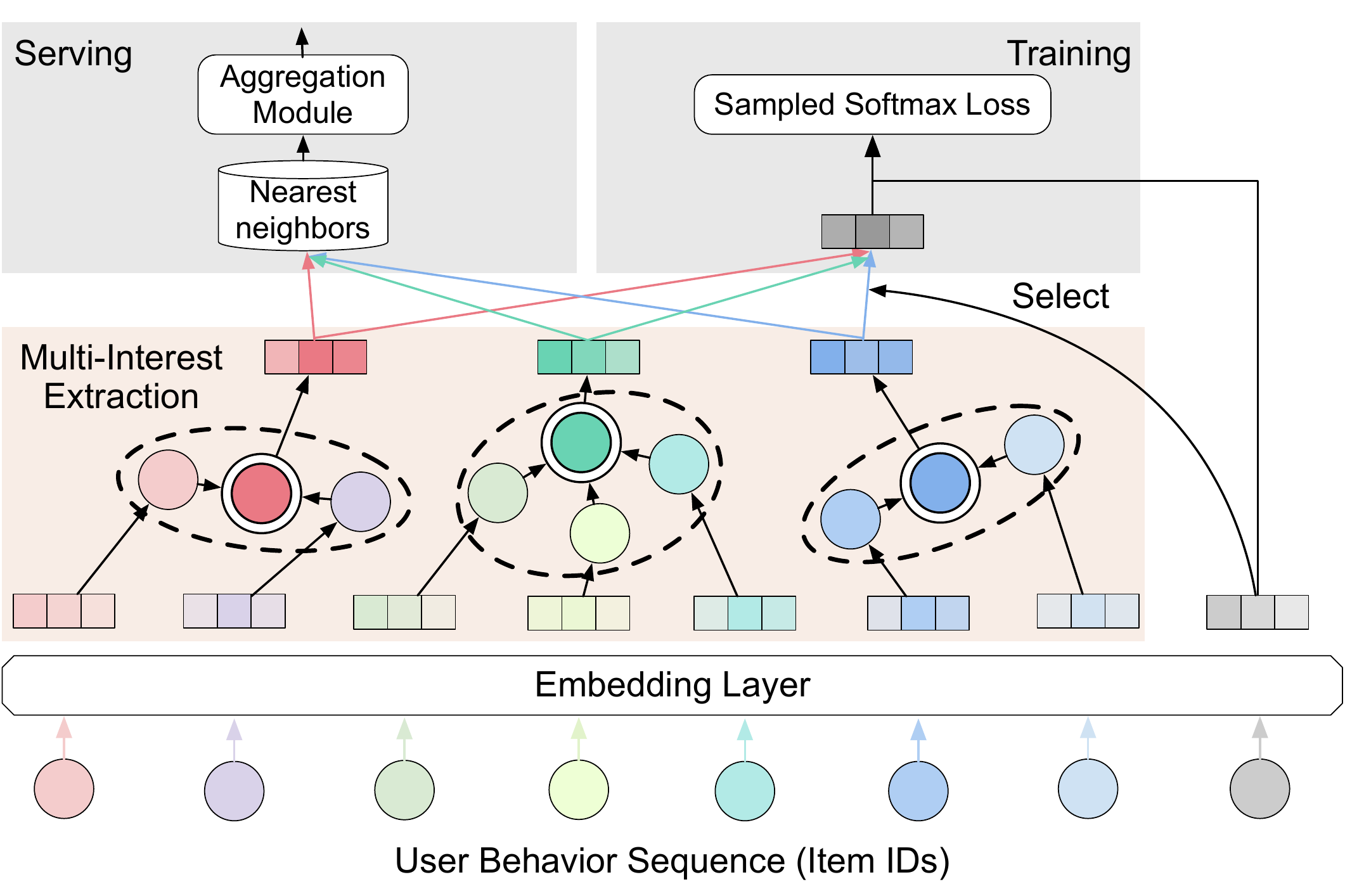}
    \caption{An overview of our model for the sequential recommendation. The input of our model is a user behavior sequence, which contains a list of item IDs. The item IDs are fed into the embedding layer and transformed into the item embeddings. Interest embeddings are generated through the multi-interest extraction module and can be then used for model training and serving. For model training, the nearest interest embedding to the target embedding will be chosen to compute the sampled softmax loss. For serving, each interest embedding will independently retrieve top-N nearest items, which are then fed into the aggregation module. The aggregation module generates the overall top-N items by a controllable procedure that balances the recommendation accuracy and diversity. }
    \label{fig:capsule_matching}
\end{figure*}

\subsection{Multi-Interest Framework}
As the item pools of industrial recommender systems usually consist of millions or even billions of items, the matching stage plays a crucial role in recommender systems. Specifically, the matching model first computes user embeddings from user historical behaviors and then retrieves a candidate set of items for each user based on the user embedding. With the help of fast K nearest neighbors (KNN) algorithm to select the closest items from the large-scale item pool to generate a candidate set for each user, we mainly focus on the computation of user embeddings. In other words, the decisive factor for the matching stage is the quality of user embeddings computed from user historical behaviors. 

Existing matching models usually use RNN\cite{hidasi2015session,wu2017recurrent} to compute embeddings for users, but most of them only generate a single embedding vector for each user. This suffers from the lack of expressiveness of a single embedding since real-world customers usually have several kinds of items in their minds and these items are often for different uses and vary a lot in categories. Such behaviors of real-world customers highlight the need to use multiple vectors to represent their multiple interests. Based on the observations, we propose a multi-interest framework for the sequential recommendation. The input of our framework is a user behavior sequence, which contains a list of item IDs representing the user's interactions with items in time order. The item IDs are fed into an embedding layer and transformed into item embeddings. A multi-interest extraction module receives item embeddings and generates multiple interests for each user.


To build a multi-interest extraction module, there are many optional methods. In this paper, we explore two methods, dynamic routing method and self-attentive method, as our multi-interest extraction module. Our framework using a dynamic routing method or self-attentive method is named as \model-DR or \model-SA, respectively. 

\vpara{Dynamic Routing.}
We utilize a dynamic routing method as a multi-interest extraction module for user behavior sequences. The item embeddings of the user sequence can be viewed as primary capsules, and the multiple user interests can be seen as interest capsules. We use the dynamic routing method from CapsNet~\cite{sabour2017dynamic}. 
We briefly introduce dynamic routing for computing vector inputs and outputs of capsules. A capsule is a group of neurons whose activity vectors represent the instantiation parameters of a specific type of entity such as an object or an object part~\cite{sabour2017dynamic}. The length of the output vector of a capsule represents the probability that the entity represented by the capsule is in the current input. Let $\mathbf{e}_i$ be the capsule $i$ of the primary layer. We then give the computation of the capsule $j$ of the next layer based on primary capsules. 
We first compute the prediction vector as
\begin{equation}
    \hat{\mathbf{e}}_{j|i}=\mathbf{W}_{ij} \mathbf{e}_{i},
\end{equation}
where $\mathbf{W}_{ij}$ is a transformation matrix. Then the total input to the capsule $j$ is the weighted sum over all prediction vectors $\hat{\mathbf{e}}_{j|i}$ as
\begin{equation}
        \mathbf{s}_j = \sum_i c_{ij} \hat{\mathbf{e}}_{j|i},
\end{equation}
where $c_{ij}$ are the coupling coefficients that are determined by the iterative dynamic routing process. The coupling coefficients between capsule $i$ and all the capsules in the next layer should sum to 1. We use ``routing softmax'' to calculate the coupling coefficients using initial logits $b_{ij}$ as
\begin{equation}
    c_{ij}=\frac{\exp(b_{ij})}{\sum_k \exp(b_{ik})},
\end{equation}
where $b_{ij}$ represents the log prior probability that capsule $i$ should be coupled to capsule $j$. A non-linear "squashing" function~\cite{sabour2017dynamic} is proposed to ensure short vectors to get shrunk to almost zero length and long vectors to get shrunk to a length slightly below 1. Then the vector of of capsule $j$ is computed by
\begin{equation}
    \label{eqn:squash}
    \mathbf{v}_j = \operatorname{squash}(\mathbf{s}_j) =  \frac{\|\mathbf{s}_j\|^2}{1+\|\mathbf{s}_j\|^2} \frac{\mathbf{s}_j}{\|\mathbf{s}_j\|},
\end{equation}
where $\mathbf{s}_j$ is the total input of capsule $j$. To calculate the output capsules $\mathbf{v}_j$, we need to calculate the probability distribution based on the inner production of $\mathbf{v}_j$ and $\mathbf{e}_i$. The calculation of $\mathbf{v}_j$ relies on itself; thus, dynamic routing method is proposed to solve this problem. The whole dynamic routing process is listed in Algorithm~\ref{algo:dynamic_routing}. The output interest capsules of the user $u$ are then formed as a matrix $\mathbf{V}_u=[\mathbf{v}_1, ..., \mathbf{v}_K] \in \mathbb{R}^{d\times K}$ for downstream tasks.



\begin{algorithm}[t]
	\caption{Dynamic Routing \label{algo:dynamic_routing}}
	\KwIn{primary capsules $\mathbf{e}_i$, iteration times $r$, number of interest capsules $K$}
	\KwOut{interest capsules $\{\mathbf{v}_j, j=1,...,K\}$}
	for each primary capsule $i$ and interest capsule $j$: initialize $b_{ij} = 0$. \\
    \For{$iter = 1,\cdots,r$} {
        for each primary capsule $i$: $\mathbf{c}_i = \operatorname{softmax}(\mathbf{b}_{i})$.\\
        for each interest capsule $j$: $\mathbf{s}_j = \sum_{i} c_{ij}\mathbf{W}_{ij}\mathbf{e}_i$.\\

        for each interest capsule $j$: $\mathbf{v}_j = \operatorname{squash}(\mathbf{s}_j)$. \\

        for each primary capsule $i$ and interest capsule $j$: $b_{ij} = b_{ij}+ \mathbf{v}_j^\top \mathbf{W}_{ij}\mathbf{e}_i$.
    }
    \Return{$\{\mathbf{v}_j, j=1,...,K\}$}
\end{algorithm}

\vpara{Self-Attentive Method.} 
The self-attentive method~\cite{lin2017structured} can also be applied to our multi-interest extraction module. 
Given the embeddings of user behaviors, $\mathbf{H}\in \mathbb{R}^{d\times n}$, where $n$ is the length of the user sequence, we use the self-attention mechanism to obtain a vector of weights $\mathbf{a} \in \mathbb{R}^{n}$:
\begin{equation}
    \mathbf{a} = \operatorname{softmax}(\mathbf{w}_{2}^\top \operatorname{tanh}(\mathbf{W}_{1} \mathbf{H}))^\top,
\end{equation}
\noindent where $\mathbf{w}_{2}$ and $\mathbf{W}_{1}$ are trainable parameters with size $d_a$ and $d_a \times d$, respectively. The superscript $\top$ denotes the transpose of the vector or the matrix. The vector $\mathbf{a}$ with size $n$ represents the attention weight of user behaviors. When we sum up the embeddings of user behaviors according to the attention weight, we can obtain a vector representation $\mathbf{v}_u = \mathbf{H} \mathbf{a}$ for the user. For the self-attentive method to make use of the order of user sequences, we add trainable positional embeddings~\cite{vaswani2017attention} to the input embeddings. The positional embeddings have the same dimension $d$ as the item embeddings and the two can be directly summed. 

This vector representation focuses on and reflects a specific interest of the user $u$. To represent the overall interests of the user, we need multiple $\mathbf{v}_u$ from the user behaviors that focus on different interests. Thus we need to perform multiple times of attention. We extend the $\mathbf{w}_{2}$ into a $d_a$-by-$K$ matrix as $\mathbf{W}_{2}$. Then the attention vector $\mathbf{a}$ becomes an attention matrix $\mathbf{A}$ as
\begin{equation}
    \mathbf{A} = \operatorname{softmax}(\mathbf{W}_{2}^\top \operatorname{tanh}(\mathbf{W}_{1} \mathbf{H}))^\top.
\end{equation}

The final matrix of user interests $\mathbf{V}_u$ can be computed by
\begin{equation}
    \label{eqn:sa}
    \mathbf{V}_u = \mathbf{H} \mathbf{A}.
\end{equation}

\vpara{Model Training.}
After computing the interest embeddings from user behaviors through the multi-interest extraction module, we use an \textit{argmax} operator to choose a corresponding user embedding vector for a target item $i$:

\begin{equation}
    \label{eqn:argmax}
    \begin{aligned}
        \mathbf{v}_u = \mathbf{V}_u[:, \operatorname{argmax}(\mathbf{V}_u^\top \mathbf{e}_i)],
    \end{aligned}
\end{equation}
where $\mathbf{e}_i$ denotes the embedding of the target item $i$, and $\mathbf{V}_u$ is the matrix formed by user interest embeddings. 

Given a training sample $(u,i)$ with the user embedding $\mathbf{v}_u$ and the item embedding $\mathbf{e}_i$, we can compute the likelihood of the user $u$ interacting with the item $i$ as

\begin{equation}
    \label{eqn:likelihood}
    P_\theta(i|u) = \frac{\exp(\mathbf{v}_u^\top \mathbf{e}_i)}{\sum_{k\in\mathcal{I}}\exp(\mathbf{v}_u^\top \mathbf{e}_k)}.
\end{equation}

The objective function of our model is to minimize the following negative log-likelihood

\begin{equation}
    \label{eqn:loss}
    loss = \sum_{u\in \mathcal{U}} \sum_{i\in \mathcal{I}_u} -\log P_\theta(i|u).
\end{equation}

The sum operator of equation (\ref{eqn:likelihood}) is computationally expensive; thus, we use a sampled softmax technique~\cite{jean2014using, covington2016deep} to train our model.

\vpara{Online Serving.}
For online serving, we use our multi-interest extraction module to compute multiple interests for each user. Each interest vector of a user can independently retrieve top-N items from the large-scale item pool by the nearest neighbor library such as Faiss~\cite{JDH17}. The items retrieved by multiple interests are fed into an aggregation module to determine the overall item candidates. Finally, the items with higher ranking scores will be recommended for users.


\begin{algorithm}[t]
	\caption{Greedy Inference \label{algo:greedy_infer}}
	\KwIn{Candidate item set $\mathcal{M}$, number of output items $N$}
	\KwOut{Output item set $\mathcal{S}$}
	$\mathcal{S} = \varnothing$ \\
    \For{$iter = 1,\cdots,N$} {
        $j = \operatorname{argmax}_{i \in \mathcal{M} \backslash \mathcal{S}} \left( f(u, i) + \lambda \sum_{k \in \mathcal{S}} g(i,k) \right)$ \\
        $\mathcal{S} = \mathcal{S} \cup \{j\}$
    }
    \Return{$\mathcal{S}$}
\end{algorithm}

\subsection{Aggregation Module}
After the multi-interest extraction module, we obtain multiple interest embeddings for each user based on his/her past behavior. Each interest embedding can independently retrieve top-N items based on the inner production proximity. But how to aggregate these items from different interests to obtain the overall top-N items? A basic and straightforward way is to merge and filter the items based on their inner production proximity with user interests, which can be formalized as
\begin{equation}
    f(u,i) = \max_{1\leq k\leq K}(\mathbf{e}_i^\top \mathbf{v}_u^{(k)}),
\end{equation}
where $\mathbf{v}_u^{(k)}$ is the $k$-th interest embedding of the user $u$. This is an effective method for the aggregation process to maximize the recommendation accuracy. However, it is not all about the accuracy of current recommender systems. People are more likely to be recommended with something new or something diverse. 
The problem can be formulated in the following. Given a set $\mathcal{M}$ with $K\cdot N$ items retrieved from $K$ interests of a user $u$, find a set $\mathcal{S}$ with $N$ items such that a pre-defined value function is maximized. Our framework uses a controllable procedure to solve this problem. We use the following value function $Q(u,S)$ to balance the accuracy and diversity of the recommendation by a controllable factor $\lambda \geq 0$,
\begin{equation}
    Q(u,\mathcal{S}) = \sum_{i\in \mathcal{S}} f(u,i) + \lambda \sum_{i\in \mathcal{S}} \sum_{j\in \mathcal{S}} g(i,j).
\end{equation}
\noindent Here $g(i,j)$ is a diversity or dissimilarity function such as
\begin{equation}
    g(i,j) = \delta(\operatorname{CATE}(i) \neq \operatorname{CATE}(j)).
\end{equation}
where $\operatorname{CATE}(i)$ means the category of item $i$ and $\delta(\cdot)$ is an indicator function. 
For the most accurate case, i.e., $\lambda=0$, we just use the above straightforward method to obtain the overall items. For the most diverse case, i.e., $\lambda=\infty$, the controllable module finds the most diverse items for users. We study the controllable factor in the Section~\ref{sec:control_study}. We propose a greedy inference algorithm to approximately maximize the value function $Q(u,S)$, which is listed in the Algorithm~\ref{algo:greedy_infer}.

\subsection{Connections with Existing Models}
We make a comparison between our model and existing models. 

\vpara{MIMN.}
MIMN~\cite{pi2019practice}, a recent representative work for the ranking stage of recommendation, uses memory networks to capture user interests from long sequential behavior data. Both MIMN and our model target at the multiple interests of users. For very long sequential behaviors, a memory-based architecture may also be insufficient to capture the long-term interests of users. Compared with MIMN, our model utilizes the multi-interest extraction module to leverage multiple interests of users instead of a complicated memory network with memory utilization regularization and memory induction unit. 

\vpara{MIND.}
MIND~\cite{li2019multi}, a recent representative work for the matching stage of recommendation, proposes a Behavior-to-Interest (B2I) dynamic routing for adaptively aggregating user's behaviors into interest representation vectors. Compared with MIND, \model-DR follows the original dynamic routing method used by CapsNet~\cite{sabour2017dynamic}, which can capture the sequential information of user behaviors. Our framework also explores a self-attentive method for multi-interest extraction. Moreover, our framework utilizes a controllable aggregation module to balance the recommendation accuracy and diversity based on users' multiple interests. 


\section{Experiments}\label{sec:exp}

In this section, we experiment on the sequential recommendation to evaluate the performance of our framework compared with other state-of-the-art methods. Besides, we also report the experimental results of our framework on a billion-scale industrial dataset. 

\begin{table}
  \centering
  \caption{\label{tab:match_stats} Statistics of datasets.}
  \begin{tabular}{c|c|c|c}
    \hline \hline
    \textbf{Dataset} & \# users & \# items & \ \# interactions \\
    \hline
    Amazon Books & 459,133 & 313,966 & 8,898,041 \\
    Taobao & 976,779 & 1,708,530 & 85,384,110 \\
    \hline \hline
  \end{tabular}
\end{table}

\begin{table*}[]
    \centering
    \caption{Model performance on public datasets. Bolded numbers are the best performance of each column. All the numbers in the table are percentage numbers with `\%' omitted.}
    \begin{tabular}{l|cccccc|cccccc}
        \hline \hline
        & \multicolumn{6}{c|}{Amazon Books} & \multicolumn{6}{c}{Taobao} \\
        & \multicolumn{3}{c}{Metrics@20} & \multicolumn{3}{c|}{Metrics@50} & \multicolumn{3}{c}{Metrics@20} & \multicolumn{3}{c}{Metrics@50} \\
        \hline
        & Recall & NDCG & Hit Rate & Recall & NDCG & Hit Rate & Recall & NDCG & Hit Rate & Recall & NDCG & Hit Rate \\
        \hline
        MostPopular & 1.368 & 2.259 & 3.020 & 2.400 & 3.936 & 5.226 & 0.395 & 2.065 & 5.424 & 0.735 & 3.603 & 9.309 \\
        YouTube DNN & 4.567 & 7.670 & 10.285 & 7.312 & 12.075 & 15.894 & 4.205 & 14.511 & 28.785 & 6.172 & 20.248 & 39.108 \\
        GRU4Rec & 4.057 & 6.803 & 8.945 & 6.501 & 10.369 & 13.666 & 5.884 & 22.095 & 35.745 & 8.494 & 29.396 & 46.068 \\
        MIND & 4.862 & 7.933 & 10.618 & 7.638 & 12.230 & 16.145 & 6.281 & 20.394 & 38.119 & 8.155 & 25.069 & 45.846 \\
        \hline
        \model-SA & \textbf{5.489} & 8.991 & 11.402 & \textbf{8.467} & \textbf{13.563} & 17.202 & \textbf{6.900} & \textbf{24.682} & 41.549 & 9.462 & 31.278 & 51.064 \\
        \model-DR & 5.311 & \textbf{9.185} & \textbf{12.005} & 8.106 & 13.520 & \textbf{17.583} & 6.890 & 24.007 & \textbf{41.746} & \textbf{9.818} & \textbf{31.365} & \textbf{52.418} \\
        \hline \hline
    \end{tabular}
    \label{tab:match_results}
\end{table*}

\subsection{Experimental Setup}


We evaluate the performance of all methods under strong generalization ~\cite{marlin2004collaborative, liang2018variational, ma2019learning}: We split all users into training/validation/test sets by the proportion of 8:1:1. We train models using the entire click sequences of training users. To evaluate, we take the first 80\% of the user behaviors from validation and test users to infer user embeddings from trained models and compute metrics by predicting the remaining 20\% user behaviors. This setting is more difficult than weak generalization where the users' behavior sequences are used during both training and evaluation processes~\cite{liang2018variational}.
In detail, we adopt a common setting of training sequential recommendation models. Let the behavior sequence of user $u$ be $(e_1^{(u)}, e_2^{(u)}, ..., e_k^{(u)}, ..., e_n^{(u)})$. Each training sample uses the first $k$ behaviors of $u$ to predict the $(k+1)$-th behavior, where $k=1,2,...,(n-1)$.

\vpara{Datasets.} We conduct experiments on two challenging public datasets. The statistics of the two datasets are shown in Table~\ref{tab:match_stats}. 

\begin{itemize}
    \item \textbf{Amazon}\footnote{\url{http://jmcauley.ucsd.edu/data/amazon/}} consists of product reviews and metadata from Amazon ~\cite{mcauley2015image,he2016ups}. In our experiment, we use the \textit{Books} category of the Amazon dataset. Each training sample is truncated at length 20.
    \item \textbf{Taobao}\footnote{\url{https://tianchi.aliyun.com/dataset/dataDetail?dataId=649\&userId=1}} collects user behaviors from Taobao's recommender systems~\cite{zhu2018learning}.  In our experiment, we only use the click behaviors and sort the behaviors from one user by time. Each training sample is truncated at length 50.
\end{itemize}

\vpara{Competitors.} We compare our proposed models, \model-SA and \model-DR, with state-of-the-art models. In our experimental setting, models should give the prediction for the unseen users of validation and test sets. Thus factorization-based methods are inappropriate for this setting.
\begin{itemize}
    \item \textbf{MostPopular} is a traditional recommendation method that recommends the most popular items to users. 
    \item \textbf{YouTube DNN}~\cite{covington2016deep} is one of the most successful deep learning models for industrial recommender systems.
    \item \textbf{GRU4Rec}~\cite{hidasi2015session} is the first work that introduces recurrent neural networks for the recommendation. 
    \item \textbf{MIND}~\cite{li2019multi} is a recent state-of-the-art model related with our model. It designs a multi-interest extractor layer based on the capsule routing mechanism, which is applicable for clustering past behaviors and extracting diverse interests.
\end{itemize}

\vpara{Implementation Notes.}
The code used by our experiments is implemented with TensorFlow\footnote{\url{https://www.tensorflow.org/}} 1.14 in Python 3.6. 

\vpara{Parameter Configuration.}
The number of dimensions $d$ for embeddings is set to 64. The number of samples for sampled softmax loss is set to 10. The number of maximum training iterations is set to 1 million. The number of interest embeddings for multi-interest models is set to 4. We use Adam optimizer~\cite{kingma2014adam} with learning rate $lr=0.001$ for optimization.


\vpara{Evaluation Metrics.}
We use the following metrics to evaluate the performance of our proposed model. We use three commonly used evaluation criteria in our experiments. 
\begin{itemize}
    \item \textbf{Recall}. We adopt per-user average instead of global average for better interpretability~\cite{karypis2001evaluation,chen2018sequential}.
        \begin{equation}
            \operatorname{Recall@N} = \frac{1}{|\mathcal{U}|}\sum_{u\in \mathcal{U}} \frac{|\mathcal{\hat{I}}_{u,N} \cap \mathcal{I}_{u}|}{|\mathcal{I}_u|},
        \end{equation}
        where $\mathcal{\hat{I}}_{u,N}$ denotes the set of top-N recommended items for user $u$ and $\mathcal{I}_u$ is the set of testing items for user $u$.
    
    \item \textbf{Hit Rate}. Hit rate (HR) measures the percentage that recommended items contain at least one correct item interacted by the user, which has been widely used in previous works~\cite{karypis2001evaluation,chen2018sequential}.
        \begin{equation}
            \operatorname{HR@N} = \frac{1}{|\mathcal{U}|}\sum_{u\in \mathcal{U}} \delta(|\mathcal{\hat{I}}_{u,N} \cap \mathcal{I}_{u}|>0),
        \end{equation}
        where $\delta(\cdot)$ is the indicator function.


    \item \textbf{Normalized Discounted Cumulative Gain}. Normalized Discounted Cumulative Gain (NDCG) takes the positions of correct recommended items into consideration~\cite{jarvelin2000ir}.
        \begin{equation}
            \operatorname{NDCG@N} = \frac{1}{Z}\operatorname{DCG@N} = \frac{1}{Z} \frac{1}{|\mathcal{U}|}\sum_{u\in\mathcal{U}}\sum_{k=1}^{N}\frac{\delta(\hat{i}_{u,k}\in \mathcal{I}_{u})}{\operatorname{log}_2(k+1)},
        \end{equation}
        where $\hat{i}_{u,k}$ denotes the $k$-th recommended item for the user $u$, and $Z$ is a normalization constant denoting the ideal discounted cumulative gain (IDCG@N), which is the maximum possible value of DCG@N. 
\end{itemize}

\subsection{Quantitative Results}
To make a fair comparison with other models, we set $\lambda=0$ in our aggregation module. We give a detailed illustration of retrieving top-N items of our framework. For our framework, each interest of a user independently retrieves top-N candidate items. Thus, our model retrieves a total of $K\cdot N$ items for each user. We sort the items by the inner product of the item embedding and the corresponding interest embedding. After the sorting, top-N items from these $K\cdot N$ items are viewed as the final candidate items of our model. The way of retrieving candidate items is also applied to MIND. 
The model performance for the sequential recommendation is shown in Table~\ref{tab:match_results}. Our models outperform all state-of-the-art models by a wide margin on all the evaluation criteria. GRU4Rec obtains the best performance over other models that only output single embedding for each user. Compared with MIND, \model-DR obtains better performance due to the difference of the dynamic routing method. \model-SA shows the strong ability to capture user interests by the self-attention mechanism and gets comparable results with \model-DR.


\begin{table}
  \centering
  \caption{\label{tab:match_ps} Model performance of parameter sensitivity. All the numbers are percentage numbers with `\%' omitted.}
  \begin{tabular}{l|cc|cc}
    \hline \hline
    & \multicolumn{2}{c|}{Amazon Books} & \multicolumn{2}{c}{Taobao} \\
    Metric@50 & Recall & NDCG & Recall & NDCG \\
    \hline
    \model-SA (K=2) & 8.835 & \textbf{14.273} & 9.935 & 32.873 \\
    \model-SA (K=4) & 8.467 & 13.563 & 9.462 & 31.278 \\
    \model-SA (K=6) & \textbf{8.901} & 14.167 & 9.378 & 31.020 \\
    \model-SA (K=8) & 8.547 & 13.631 & 9.493 & 31.196 \\
    \hline
    \model-DR (K=2) & 7.081 & 12.068 & 9.293 & 30.735 \\
    \model-DR (K=4) & 8.106 & 13.520 & 9.818 & 31.365 \\
    \model-DR (K=6) & 7.904 & 13.219 & 10.836 & \textbf{34.048} \\
    \model-DR (K=8) & 7.760 & 12.900 & \textbf{10.841} & 33.895 \\
    \hline \hline
  \end{tabular}
\end{table}

\vpara{Parameter Sensitivity.}
We investigate the sensitivity of the number of interests $K$ of our framework. Table~\ref{tab:match_ps} illustrates the performance of our framework when the hyperparameter $K$ changes. Our two models show the different properties of this hyperparameter. For the Amazon dataset, \model-SA obtains the better performance when $K=2, 6$ and \model-DR gets the best result when $K=4$. For the Taobao dataset, \model-DR gets better performance when $K$ increases from 2 to 8 but \model-SA obtains the best result when $K=2$.

\begin{table}
  \centering
  \caption{\label{tab:control} Model performance of Amazon dataset for the controllable study. All the numbers are percentage numbers with `\%' omitted.}
  \begin{tabular}{c|cc|cc}
    \hline \hline
    & \multicolumn{2}{c|}{\model-SA (K=4)} & \multicolumn{2}{c}{\model-DR (K=4)} \\
    Metric@50 & Recall & Diversity & Recall & Diversity \\
    \hline
    $\lambda=0.00$ & \textbf{8.467} & 23.237 & \textbf{8.106} & 19.036 \\
    $\lambda=0.05$ & 8.347 & 38.808 & 7.931 & 42.915 \\
    $\lambda=0.10$ & 8.229 & 46.731 & 7.850 & 46.258 \\
    $\lambda=0.15$ & 8.142 & 51.135 & 7.820 & 46.912 \\
    $\lambda=0.20$ & 8.086 & 53.671 & 7.783 & 47.581 \\
    $\lambda=0.25$ & 8.034 & \textbf{55.100} & 7.764 & \textbf{48.375} \\
    \hline \hline
  \end{tabular}
\end{table}

\subsection{Controllable Study}
\label{sec:control_study}
To obtain the final top-N candidate items for each user, we propose a novel module to aggregate the items retrieved by different interests of each user. 
In addition to aim at achieving high prediction accuracy for the recommendation, some studies suggest the need for diversified recommendations to avoid monotony and improve customers' experience~\cite{gogna2017balancing,cheng2017learning}. 

Recommendation diversity plays a more important role in current recommender systems. Many pieces of research target on improving the recommendation diversity~\cite{bradley2001improving,qin2013promoting}. Our proposed aggregation module can control the balance of recommendation accuracy and diversity. We use the following definition of individual diversity based on item categories:
\begin{equation}
    \operatorname{Diversity@N} = \frac{\sum_{j=1}^N \sum_{k=j+1}^N \delta(\operatorname{CATE}(\hat{i}_{u,j}) \neq \operatorname{CATE}(\hat{i}_{u,k}))}{N \times (N-1) / 2},
\end{equation}
\noindent where $\operatorname{CATE}(i)$ is the category of item $i$, $\hat{i}_{u,j}$ denotes the $j$-th recommended item for the user $u$, and $\delta(\cdot)$ is an indicator function. 

Table~\ref{tab:control} shows the model performance of the Amazon dataset when we control the factor $\lambda$ to balance the recommendation quality and diversity. From the table, recommendation diversity increases substantially and recall decreases slightly when the controllable factor $\lambda$ increases. Our aggregation module can achieve the optimum trade-off between the accuracy and diversity by choosing an appropriate value for the hyperparameter $\lambda$.

\subsection{Industrial Results}

\begin{table}
  \centering
  \caption{\label{tab:industrial_stats} Statistics of the industrial dataset}
  \begin{tabular}{c|c|c|c}
    \hline \hline
    \textbf{Dataset} & \# users & \# items & \# interactions \\
    \hline
    Industrial & 145,606,322 & 22,554,170 & 4,322,505,616 \\
    \hline \hline
  \end{tabular}
\end{table}



We further experiment on the industrial dataset collected by Mobile Taobao App on February 8th, 2020. The statistics of the industrial dataset are shown in the Table~\ref{tab:industrial_stats}. The industrial dataset contains 22 million high-quality items, 145 million users, and 4 billion behaviors between them. 

Our framework has been deployed on the Alibaba distributed cloud platform, where every two workers share an NVIDIA Tesla P100 GPU with 16GB memory. We split the users and use the click sequences of training users to train our model. To evaluate, we use our model to compute multiple interests for each user in the test set. Each interest vector of a user independently retrieves top-N items from the large-scale item pool by a fast nearest neighbor method. The items retrieved by different user interests are fed into our aggregation module. After this module, top-N items out of $K\cdot N$ items are the final candidate items and are used to compute the evaluation metric, recall@50. 

We conduct an offline experiment between our framework and the state-of-the-art sequential recommendation method, MIND~\cite{li2019multi}, which has shown significant improvement in the recommender system of Alibaba Group. The experimental result demonstrates that our \model-SA and \model-DR improve recall@50 by 1.39\% and 8.65\% compared with MIND, respectively.

\begin{figure}
    \centering
    \includegraphics[width=0.47\textwidth]{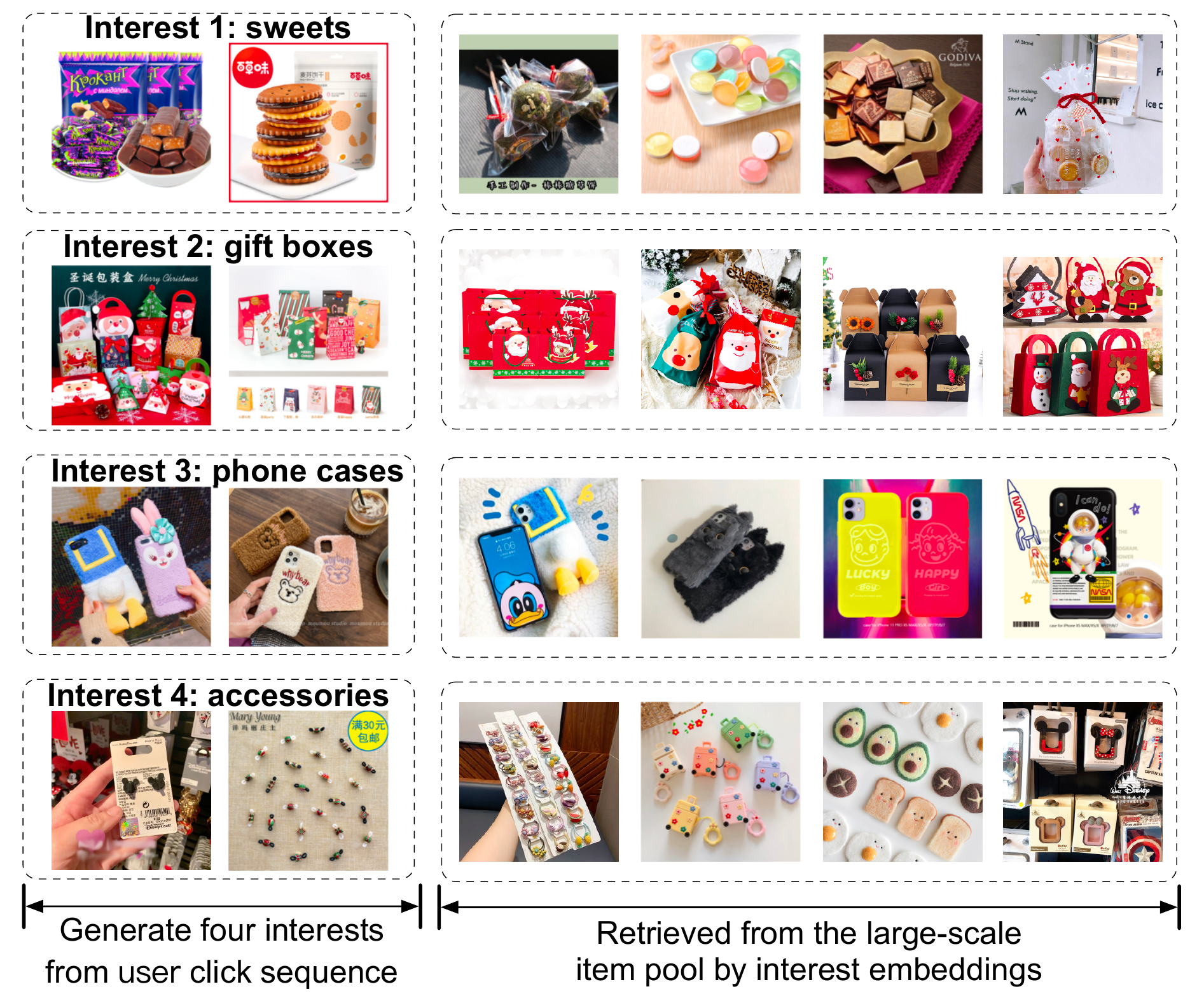}
    \caption{A case study of an e-commerce user. We generate four interest embeddings from the click sequence of a user by our model. We find that the four interests of the user are about sweets, gift boxes, phone cases, and accessories. We report those items in the click sequence that correspond to the four interests. The right part shows the items retrieved from the industrial item pool by interest embeddings. }
    \label{fig:case_study_interests}
\end{figure}

\vpara{Case Study.}
From the Figure~\ref{fig:case_study_interests}, we can see that our model learns four different interests of the user from her click sequence. It is worth noting that our model only uses item IDs for training and does not use the manually defined category information of items. Despite that, our model still can learn the item categories from user behavior sequences. Each interest learned by our model approximately corresponds to one specific category and can retrieve similar items of the same category from the large-scale industrial item pool.


\section{Conclusion} \label{sec:conclusion}

In this paper, we propose a novel controllable multi-interest framework for the sequential recommendation. Our framework uses a multi-interest extraction module to generate multiple user interests and uses an aggregation module to obtain the overall top-N items. 
Experimental results demonstrate that our models can achieve significant improvements over start-of-the-art models on two challenging datasets. Our framework has also been successfully deployed on the Alibaba distributed cloud platform. Results on the billion-scale industrial dataset further confirm the effectiveness and efficiency of our framework in practice. Recommender systems start a new phase owing to the rapid development of deep learning. Traditional recommendation methods cannot meet the requirements of the industry. For the future, we plan to leverage memory networks to capture the evolving interests of users and introduce cognitive theory to make better user modeling.

\begin{acks}
The work is supported by the
NSFC for Distinguished Young Scholar 
(61825602),
NSFC (61836013), 
and a research fund supported by Alibaba Group.
\end{acks}

\bibliographystyle{ACM-Reference-Format}
\bibliography{reference}

\appendix
\section{Appendix}
\label{sec:appendix}

In the appendix, we give the implementation notes of our proposed models. The details of other models and descriptions of datasets are then given.

\subsection{Implementation Notes}

\vpara{Running Environment.}
The experiments in this paper can be divided into two parts. One is conducted on two public datasets using a single Linux server with 4 Intel(R) Xeon(R) CPU E5-2680 v4 @ 2.40GHz, 256G RAM, and 8 NVIDIA GeForce RTX 2080 Ti. The codes of our proposed models in this part are implemented with TensorFlow\footnote{\label{fn:tf}\url{https://www.tensorflow.org/}} 1.14 in Python 3.6. 
The other part is conducted on the industrial dataset using Alibaba's distributed cloud platform\footnote{\url{https://data.aliyun.com/}} which contains thousands of workers. Every two workers share an NVIDIA Tesla P100 GPU with 16GB memory. Our proposed models are implemented with TensorFlow 1.4 in Python 2.7 in this part.

\vpara{Implementation Details.}
Our codes used by a single Linux server can be split into three parts: data iterator, model training, and evaluation. For each training iteration, the data iterator selects random training users with a size of $batch\_size$. For each selected user, we randomly select an item in his/her click sequence as the training label and use the items before that item as the training sequence. 
The training part is implemented following the training loop in the Algorithm~\ref{algo:training} based on the Tensorflow 1.x APIs. Our loss function is based on \textit{tf.nn.sampled\_softmax\_loss}.
The evaluation part replies on Faiss\footnote{\url{https://github.com/facebookresearch/faiss}}, a library for efficient similarity search and clustering of dense vectors. We use the \textit{GpuIndexFlatIP} class of Faiss, which implements an exact search for the inner product on GPU. 
All model parameters are updated and optimized by stochastic gradient descent with Adam updating rule~\cite{kingma2014adam}. 
The distributed version of our proposed models is implemented based on the coding rules of Alibaba's distributed cloud platform in order to maximize the distribution efficiency. 

\vpara{Parameter Configuration.}
Our user/item embedding dimension $d$ is set to 64. The number of samples for sampled softmax loss is set to 10. The number of maximum training iterations is set to 1 million and all models use early stopping based on the Recall@50 on the validation set. The batch size for the Amazon dataset and Taobao dataset is set to 128 and 256, respectively. The number of iterations for the dynamic routing method is set to 3. The number of interest embeddings $K$ for multi-interest models is set to 4 for a fair comparison. We use the Adam optimizer~\cite{kingma2014adam} with learning rate $lr=0.001$ for optimization.

\vpara{Code and Dataset Releasing Details.}
The code of all models and our partition of the two public datasets are available\footnote{\url{https://github.com/THUDM/ComiRec}}. 

\subsection{Compared Methods}
We give the implementation details about all compared methods as follows. 

\begin{itemize}
    \item \textbf{MostPopular} is a non-personalized method that recommends the most popular items to users. This method does not need training and we implement it separately. 
    \item \textbf{YouTube DNN} is one of the most successful deep learning models for industrial recommender systems. We implement the model in our code based on the original paper.
    \item \textbf{GRU4REC} is the first work that introduces recurrent neural networks for the recommendation. We implement the model by \textit{tf.nn.rnn\_cell.GRUCell} and \textit{tf.nn.dynamic\_rnn} of TensorFlow in our code.
    \item \textbf{MIND} is a recent state-of-the-art model. We implement the model based on the original paper and an internal version of the code in Alibaba Group. 
\end{itemize}


\subsection{Datasets}
Our experiments evaluate on three datasets, including two public datasets and a billion-scale industrial dataset. For the two public datasets, we keep users and items with at least 5 behaviors. 

\begin{itemize}
    \item \textbf{Amazon}\footnote{\url{http://jmcauley.ucsd.edu/data/amazon/}} consists of product reviews and metadata from Amazon ~\cite{mcauley2015image,he2016ups}. In our experiment, we use the \textit{Books} category of the Amazon dataset. For each user $u$, we sort the reviews from the user by time, and our task is to predict whether the user will write the review for the item based on previous reviews. Each training sample is truncated at length 20.
    \item \textbf{Taobao}\footnote{\url{https://tianchi.aliyun.com/dataset/dataDetail?dataId=649\&userId=1}} collects user behaviors from Taobao's recommender systems~\cite{zhu2018learning}. Taobao dataset randomly selects about 1 million users who have behaviors including click, purchase, add-to-cart, and add-to-preference from November 25 to December 03, 2017. Each behavior is represented by five fields, which consist of user ID, item ID, item's category ID, behavior type, and timestamp. In our experiment, we only use the click behaviors and sort the behaviors from one user by time. Each training sample is truncated at length 50.
    \item \textbf{Industrial dataset} collects user click behaviors by Mobile Taobao App on February 8th, 2020. The industrial dataset contains 22 million high-quality items, 145 million users, and 4 billion behaviors between them. Each training sample is truncated at length 40.
\end{itemize}

\begin{algorithm}[t]
	\caption{\model \label{algo:training}}
	\KwIn{User behavior sequences. }
	Initialize all the model parameters. \\
	Generate training samples $\{(u,i)\}$ with user click sequences.\\
	\While{not converged} {
		\For{each batch from training samples} {
			Compute $\mathbf{V}_u$ using multi-interest extraction module. \\
			Compute $\mathbf{v}_u$ based on Equation (\ref{eqn:argmax}). \\
			Compute sampled softmax loss using Equation (\ref{eqn:loss}). \\
			Update model parameters by the Adam optimizer.
		}
	}
\end{algorithm}


\end{document}